\documentclass[a4paper,fleqn,usenatbib,useAMS]{mnras}

\usepackage{newtxtext,newtxmath}
\usepackage[T1]{fontenc}
\usepackage{ae,aecompl}

\usepackage{array}
\usepackage{multicol}
\usepackage{hyperref}
\usepackage{xcolor}
\usepackage{graphicx}	% Including figure files
\usepackage{natbib}	
\usepackage{amsmath,amsfonts,amssymb} % Advanced maths commands
\title[Modelling a compact jet using disc-driven shocks]{Modelling the compact jet in MAXI J1836-194 with disc-driven shocks}

\author[M. Péault et al.]{
M. Péault,$^{1}$\thanks{E-mail: mathias.peault@irap.omp.eu}
J. Malzac,$^{1}$
M. Coriat,$^{1}$
T.D. Russell,$^{2}$
K.I.I. Koljonen,$^{3,4,5}$
\newauthor  \ R. Belmont,$^{1,6}$
S. Corbel,$^{6,7}$
S. Drappeau,$^{1}$
J. Ferreira,$^{8}$
P.O. Petrucci,$^{8}$
\newauthor
 \ J. Rodriguez,$^{6}$
  D.M. Russell$^{5}$\\
  \\
$^{1}$IRAP, Université de Toulouse, CNRS, UPS, CNES, Toulouse, France\\
$^{2}$Anton Pannekoek Institute for Astronomy, University of Amsterdam, Science Park 904, NL-1098 XH Amsterdam, the Netherlands\\
$^{3}$Finnish Centre for Astronomy with ESO (FINCA), University of Turku, Väisäläntie 20, 21500 Piikkiö, Finland\\
$^{4}$Aalto University Metsähovi Radio Observatory, PO Box 13000, FI-00076 Aalto, Finland\\
$^{5}$New York University Abu Dhabi, PO Box 129188, Abu Dhabi, UAE\\
$^{6}$Laboratoire AIM (CEA/IRFU - CNRS/INSU - Université Paris Diderot), CEA DRF/IRFU/DAp, F-91191 Gif-sur-Yvette, France\\
$^{7}$Station de Radioastronomie de Nançay, Observatoire de Paris, PSL Research University, CNRS, Univ. Orléans, 18330 Nançay, France\\
$^{8}$Université Grenoble Alpes, CNRS, IPAG, F-38000 Grenoble, France
}

\date{Accepted XXX. Received YYY; in original form ZZZ}

\pubyear{2017}

\begin{document}
\label{firstpage}
\pagerange{\pageref{firstpage}--\pageref{lastpage}}
\maketitle

\begin{abstract}
The black hole candidate MAXI J1836-194 was discovered in 2011 when it went into an outburst, and was the subject of numerous, quasi-simultaneous, multi-wavelength observations in the radio, infrared, optical and X-rays. In this paper, we model its multi-wavelength radio to optical spectral energy distributions (SEDs) with an internal shock jet model. The jet emission is modelled on five dates of the outburst, during which the source is in the hard and hard intermediate X-ray spectral states. The model assumes that fluctuations of the jet velocity are driven by the variability in the accretion flow which is traced by the observed X-ray timing properties of the source. \textcolor{black}{While the global shape of the SED is well reproduced by this model for all the studied observations, the variations in bolometric flux and typical energies require at least two parameters to evolve during the outburst. Here we investigate variations of the jet power and mean Lorentz factor, which are both  found to increase with the source luminosity.} Our results are compatible with the evolution of the jet Lorentz factor reported in earlier studies of this source. However, due to the large degeneracy of the parameters of the \textsc{ishem} model, our proposed scenario is not unique.\\
\end{abstract}
%A similar variation of the jet Lorentz factor has already been suggested by \cite{gamup} as a possible interpretation of the very peculiar radio/X-ray correlation observed in this source.


%%%%%%%%%%%%%%%%%%%%%%%%%%%%%%%%%%%%%%%%%%%%%%%%%%

%%%%%%%%%%%%%%%%% BODY OF PAPER %%%%%%%%%%%%%%%%%%
\section{Introduction} %%%%%%% HERE THE INTRODUCTION %%%%%%%%%%%%%%%%%%%%%%%%%%%%%%%%%%%%%%%%%%
\label{intro}
\indent In the early 2000s, the study of compact jets in X-ray binaries (XRBs) boomed with the discovery of the characteristic flat -or slightly inverted- radio spectra in the hard X-ray spectral state \citep[see][]{Corbel,Fender,CF2002}. This observed flat radio emission is attributed to partially self-absorbed synchrotron emission from a jet \citep{Blan,BP}. Specifically, the flat radio spectrum results from neglecting the cooling of the electrons and, particularly, the energy losses due to the adiabatic expansion of the jet in the external medium. However, in the absence of an acceleration mechanism that continuously compensates for the adiabatic energy losses along the jet, cooling can result in a highly inverted radio spectrum, therefore inconsistent with the observations. \cite{M2013, M2014} showed that internal shocks caused by rapid fluctuations of the jet velocity constitute an effective dissipation mechanism that can release energy over a broad range of scales along the jet. In this model, the dissipation profile along the jet and the resulting shape of the SED are determined almost entirely by the power spectrum of the velocity fluctuations. The other parameters of the model (such as jet power and jet opening angle) can only shift the SED in photon frequency or in normalisation. Interestingly, the adiabatic losses are totally compensated in the case of internal shock jet models where shells of matter are ejected at the base of the jet with Lorentz factors that follow flicker noise fluctuations (i.e. the power spectral density is inversely proportional to the frequency, PSD $\propto 1/$f), maintaining the flat jet spectral slope that is usually observed. In XRBs, it turns out that the X-ray light curve which can, in principle, act as a tracer of the fluctuations in the accretion flow (or mass accretion rate), often presents a power spectrum that is close to $1/$f within a certain range of Fourier frequencies. This coincidence led \cite{M2013, M2014} to suggest that the fast fluctuations of the jet velocity causing internal shocks might be driven by the variability of the accretion flow.\\
\indent \cite{Drap} first explored this idea by using the internal shock code \textsc{ishem} \citep{M2014} to show that an observed radio-IR jet SED from the black hole binary GX 339-4 during the hard state can be well reproduced, under the assumption that the power spectrum of the jet fluctuations is identical to the fluctuations in the disc observed in X-rays. \cite{darkj} suggested that the quenching of the radio emission in the soft X-ray spectral state could be associated with the much weaker X-ray variability present in this state. Dark jets could be present in the soft state carrying a similar power as in the hard state, but weaker shocks due to the smaller amplitude of the velocity fluctuations mean the jet would be undetectable. While these results were encouraging, they need to be applied to other sources and observations at various phases of an outburst to test their universality.\\
 \indent The galactic black hole candidate MAXI J1836-194, first detected on 2011 August 30 during an outburst \citep{Negoro}, represents an interesting opportunity for modelling as it offers the possibility to work with an excellent multi-wavelength coverage ranging from radio to X-rays. Additionally, in most sources the optical emission is dominated by the accretion disc while MAXI J1836-194 appears to be jet dominated \citep{RUSSB} in most states, making it an excellent target for jet studies. \\
 \indent \cite{RUSSA} constrained the distance to MAXI J1836-194 between 4kpc and 10kpc and its disc inclination between $4\degr$ and $15\degr$. The jet dominance is most likely related to the small angle between the jet and the line of sight. The mass of the black hole was estimated between 7.5M$_{\odot}$ and 11M$_{\odot}$ based on X-ray spectral fitting using the TCAF model \citep[see][for details]{blackHmass}. {\textcolor{black}{We note that these mass estimates are model dependent. We will nevertheless use it as a guide for lack of better constraints.} \\
\indent In this paper, we model the spectral energy distributions of the compact jet of MAXI J1836-194 on five dates of its 2011 outburst with the \textsc{ishem} code, using the quasi-simultaneously observed X-ray power spectra as an input of the model. In section \ref{Obse} we present the observations and describe the main features of the model, discussing the influence of the parameters on the synthetic SEDs. In section \ref{hivps} we show our resulting SEDs along with the final parameters. {\textcolor{black}{In section \ref{paraspace}, we perform an in-depth exploration of the parameter space and identify alternative solutions to match the observations.} Finally in Section \ref{dis} we discuss our results by comparing them to conclusions drawn in previous works.
\section{Methods}%%%%%%% SECTION TWO %%%%%%%%%%%%%%%%%%%%%%%%%%%%%%%%%%%%%%%%%%
\label{Obse}
\subsection{Observations} 
The black hole candidate MAXI J1836-194 was intensively observed during its 2011 outburst by numerous instruments in different spectral bands \citep[see][and references therein]{DATADMRUSS, RUSSA}. The large amount of collected data, ranging from the radio domain to X-rays, makes this two-month period ideal for a multi-wavelength study of its compact jet. In this work, we use observations presented in \cite{DATADMRUSS} and in \cite{RUSSA}, namely, radio data collected by the Karl G. Jansky Very Large Array (VLA), \textcolor{black}{submillimeter} data obtained with the \textcolor{black}{Submillimeter} array, mid-IR data collected by the Very Large Telescope (VLT), optical observations obtained with the two 2m Faulkes Telescopes and with Swift UVOT (optical+UV) and finally X-ray data gathered by Swift X-ray telescope (XRT) and the Rossi X-ray Timing Explorer (RXTE). There were also optical+NIR observations collected with GROND \citep{grond}. All these observations were taken during a failed hard-to-soft state transition \citep{fail} that occurred between the beginning of September and the end of October \citep{ferrigno}. During this period, the source went into a hard-intermediate state (HIMS) but never reached the soft state and, instead, the outburst "failed" and went back to the hard state. \\
\indent An investigation of how the compact jet evolved during the two-month outburst was performed by \cite{DATADMRUSS, RUSSA, gamup} notably reporting the interesting behaviour of the break frequency that marks the transition between the optically thick and the optically thin part of the jet spectrum. This spectral break (also synchrotron emission peak), already detected in some XRBs \citep{breakgx,indexindex,DATADMRUSS}, corresponds to the base of the particle acceleration region in the jet \citep[see][]{region,MRKFF,Chaty}. As the source hardened during the decay phase, the break moved to higher frequencies while the optical-IR flux brightened and the radio flux faded, making the jet spectrum more and more inverted. Studies of the evolution of the compact jet in MAXI J1836-194 suggested the existence of a relation between the break frequency and the hardness \citep[see][]{RUSSA} and even suggested that this shift, along with the IR fading/brightening, could be driven by the jet quenching/recovery during the outburst \citep{RUSSA, hersch}. \cite{karri} also showed that this jet break/hardness relation appears to be a common feature in black hole X-ray binaries and low luminosity active galactic nuclei.\\
\indent We model five of the six epochs of multi-wavelength observations (taken on 2011 September 03, September 17, September 26, October 12 and October 27) which trace each step of the failed transition in order to track the jet evolution. The sixth epoch (September 12) is not studied in this paper as there was no simultaneous RXTE observations, therefore no X-ray PSD to use as input for our jet model. MAXI J1836-194 was in the hard state on September 03, then in the HIMS on September 17 and 26, and finally back in the hard state on October 12 and 27 \citep{ferrigno}.
\subsection{Jet model}
\begin{figure*}
\centering
\includegraphics[scale=0.56]{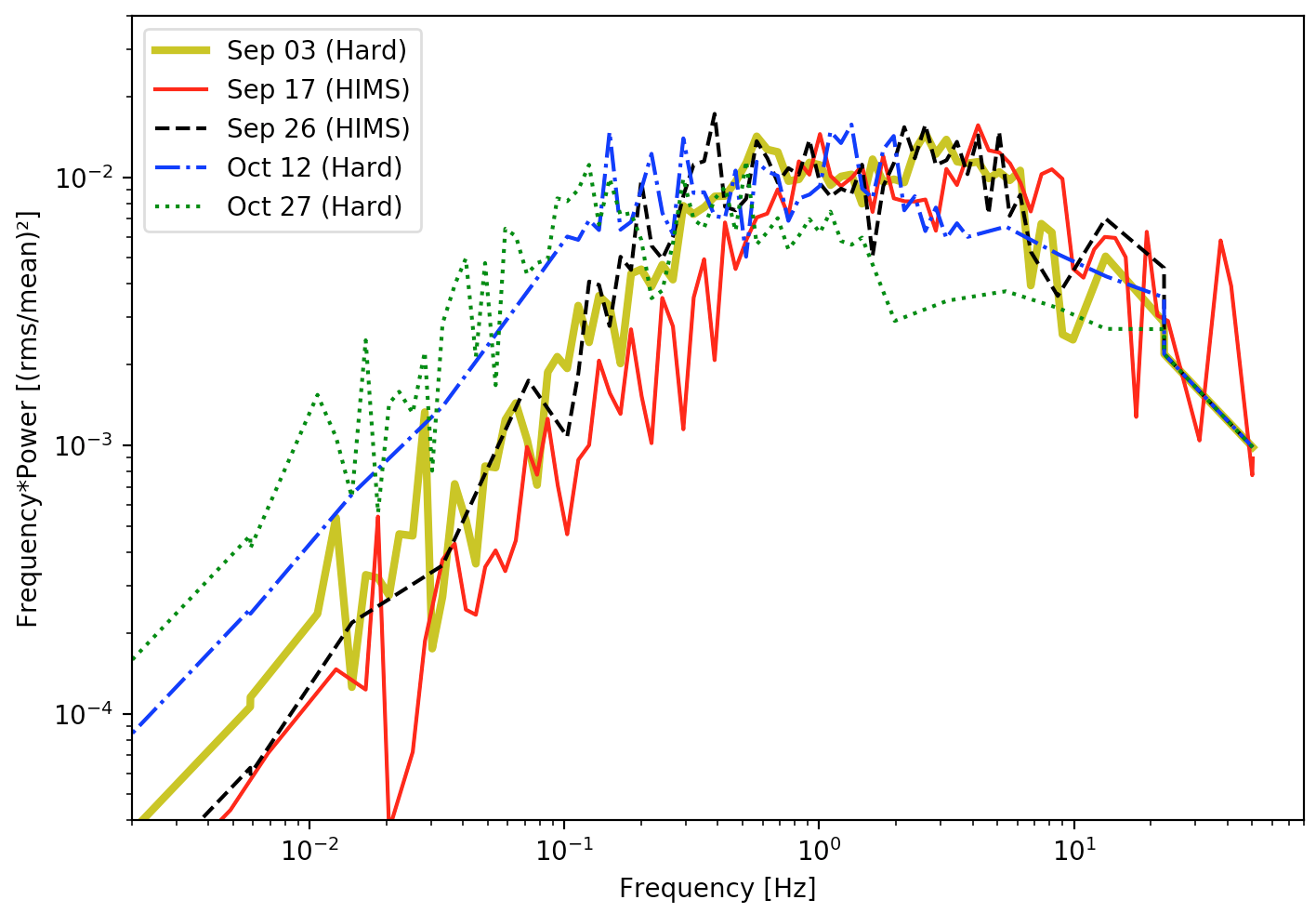}
\caption{Power spectral densities provided as input to the \textsc{ishem} code and used to simulate the ejection of the shells of matter \textcolor{black}{(plotted here in frequency*power on the y-axis for visual purposes)}.
 They have been computed using the quasi-simultaneous X-ray light curves measured for the five dates of the study with RXTE. \textcolor{black}{We observe} variations in the shape of the PSDs during the observation period, notably at low frequencies, highlighting how necessary it is to use the coinciding measure of the variability to ensure the accuracy of the outgoing jet spectrum.}
\label{Tux}
\end{figure*}
To reproduce the spectra of the compact jet in MAXI J1836-194, we used the numerical code \textsc{ishem} presented in \cite{M2014}. In this model, the emission of the jet is powered by internal shocks. These internal shocks appear when homogeneous shells of matter are ejected at the base of the jet with variable velocities \citep{M2014}. Fast ejecta catch up with slow ejecta creating shock waves that release a fraction of the bulk kinetic energy of the shells of matter into the jet and cause the acceleration of electrons. This leads to synchrotron emission and possibly inverse-Compton emission that make the jet observable (presently, only synchrotron emission is considered in \textsc{ishem}). Following \cite{Drap}, the fluctuations of the jet Lorentz factor are generated such that their power spectrum is identical to the observed X-ray power spectrum.\\
\indent To obtain information about the timing properties of the X-ray emission during the outburst of MAXI J1836-194, we used X-ray observations from the Proportional Counter Array (PCA) instrument \citep{Jahoda} onboard RXTE. We searched for X-ray observations that were taken within a day of the observations at other wavebands (the daily variations are slow in the hard state). We extracted light curves with a time bin of 2$^{-11}$ sec from the single bit mode data (SB\_125us\_0\_249\_1s) using HEASOFT 6.19. The light curves were further analyzed in ISIS \citep[Interactive Spectral Interpretation System,][]{HOUCK} using the SITAR (S-lang/ISIS Timing Analysis Routines) package to form the PSD. We calculated the PSD for every 512 sec segments rejecting those with data gaps and averaging all PSDs over the whole light curve. The resulting PSDs were binned logarithmically with $\delta$f/f= 0.1, and the Poisson noise and dead-time effects were removed by fitting a constant to the Poisson noise-dominated part of the PSD and removing the constant from the X-ray power of each frequency bin. The observed PSDs are limited to frequencies above 1/512 Hz. As \textsc{ishem} requires also information on longer time scales, we extrapolate the PSDs to lower frequencies as flat noise. \textcolor{black}{The final X-ray PSDs used as input in \textsc{ishem} and corresponding to the five observations are shown in Figure \ref{Tux}.}\\
\indent \textcolor{black}{As usually observed in X-ray binaries in the hard state, the low frequency variability is gradually suppressed as the source gets closer to softer states (the low break frequency in the PSD shifts towards higher frequencies).} In the framework of our model, this evolution of the PSD impacts the shape of the radio-IR SEDs. \textcolor{black}{For the sake of clarity, the error bars on the five PSDs are not plotted. Nevertheless, these error bars are significant and we therefore studied their impact on the resulting spectra by running several \textsc{ishem} simulations in which we randomly varied the PSDs within the confidence intervals. It results a minor impact on the SEDs, insignificantly changing the parameters values.}
 \subsection{Parameters}
\label{susu} 
\begin{figure*}
\centering
\includegraphics[scale=0.65]{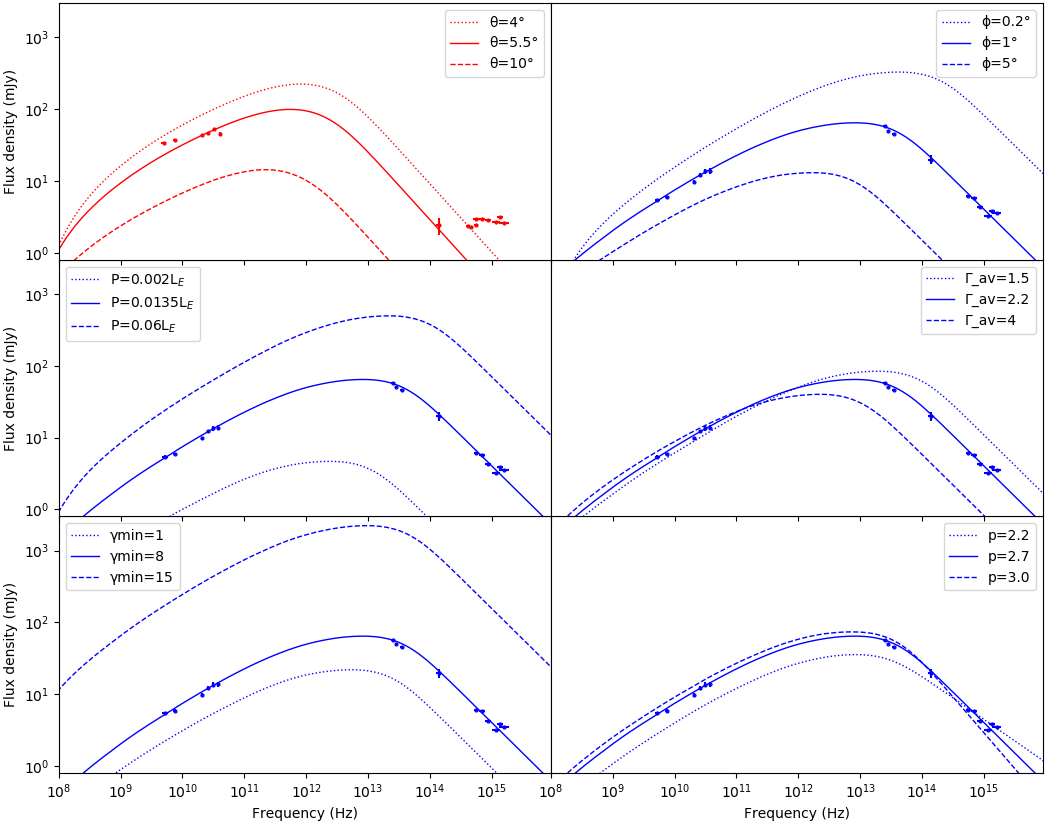}
\caption{The effect of a number of the main model parameters on the September 17 (red, top left, highest $\Gamma_{\text{av}}$) and September 26 (blue) calculated jet spectra. \textcolor{black}{On each panel, a unique parameter is altered from the parameters displayed in Table \ref{my-label} and in its caption.} (Top left) the source inclination angle. This effect is not clearly visible on September 26 due to a too low value of $\Gamma_{\text{av}}$, thus we use the epoch with the highest $\Gamma_{\text{av}}$. (Top right) the jet opening angle, (Middle left) the jet power, (Middle right) the mean Lorentz factor of the ejected shells of matter, (Bottom left) the lower limit of the electron distribution and (Bottom right) the electron distribution index. The dotted lines show a range of input values.}
\label{para}
\end{figure*}
We used three different families of parameters in our study: parameters related to the global properties of the source, parameters related to the jet itself, and parameters related to the distribution of the radiating particles. The first family consists of the distance to the source (D), the source orbital inclination with respect to the line of sight ($\theta$) and the mass of the central black hole. The second family includes the jet power (P), the jet opening angle ($\phi$) and the parameters involved in the launching of shells of matter: radius at the base of the jet, mass, bulk Lorentz factor ($\Gamma_{\text{av}}$), volume filling factor (f$_{\text{v}}$). We also include the parameters that define how the energy is liberated (sub/supersonic collisions, electron/proton equipartition) or lost (radial/longitudinal losses). Finally, in the third family, we define the parameters that characterise the distribution of the radiating particles. In these simulations, we only compute the synchrotron emission from electrons with a power-law energy distribution. We then include, in this family, the lower and upper energy limit ($\gamma_{\text{min}}$ and $\gamma_{\text{max}}$) and the index of the power law ($p$) of the electron distribution.\\ 
\indent Within these three categories, some simulation parameters are well constrained by the observations while others are not. Consequently, certain parameters are left free but confined in physically consistent intervals, while others are fixed because they have a negligible impact on the resulting SEDs. The impact of the simulation parameters on the break flux and frequency can be estimated analytically \citep[see][]{M2013, M2014}. The relevant scalings for the flux normalisation and for the position of the break frequency are as follows:
\begin{equation}
\label{flux}
\text{F$_{\nu_{\text{break}}}$} \propto\frac{\delta^{2}\: i_{\text{$\gamma$}}^{5/(p+4)}}{ D^{2}_{\text{kpc}}\text{tan}(\phi)} \left [\frac{{P}}{(\mathit{\Gamma_{\text{av}}}+1)\mathit{\Gamma_{\text{av}}} \textcolor{black}{\beta}} \right ]^{(2p+13)/(2p+8)} \ ,
\end{equation}
\begin{equation}
\label{freq}
\text{$\nu_{\text{break}}$}\propto\frac{\delta \: i_{\text{$\gamma$}}^{2/(p+4)}}{ \text{tan}(\phi) } \frac{P^{(p+6)/(2p+8)}}{ \left [(\mathit{\Gamma_{\text{av}}}+1)\mathit{\Gamma_{\text{av}}}\textcolor{black}{\beta}\right ] ^{(3p+14)/(2p+8)} }\ ,
\end{equation}
 where \text{$\beta$}=$\sqrt{1-\Gamma_{\text{av}}^{-2}}$, \text{$\delta$}=$[\Gamma_{\text{av}}(1-\beta \text{cos} \theta)]^{-1}$ and $i_{\text{$\gamma$}}$=(2-p)($\gamma_{\text{max}}^{2-p}-\gamma_{\text{min}}^{2-p})^{-1}$.
 Figure \ref{para} shows the effect of the simulation parameters on the jet spectra. \\
\indent \textcolor{black}{Regarding the first family, the parameters are not well constrained by the observations (see Section \ref{intro}). In \textsc{ishem}, the black hole mass has an impact on the jet power (simply because the jet power is expressed as a fraction of the Eddington luminosity, L$_{\text{E}}$) but also controls the initial radius of the shells of matter since the latter is expressed in gravitational radii (R$_{\text{g}}$=$\frac {\text{GM}}{\text{c$^{2}$}}$). Consequently, it only has a small effect on the final SEDs.} The distance to the source, D, has a strong effect on the resulting \textcolor{black}{flux densities} but has no impact on the location of the break frequency because it does not affect the emission mechanism. As for the inclination, $\theta$, due to relativistic beaming, a small angle leads to high fluxes and to high break frequencies. However, the shift in the break frequency becomes only noticeable for high values of bulk Lorentz factor (see Figure \ref{para}, top left panel). \\
 \indent Jets are collimated ejections of matter, therefore their opening angle, $\phi$, should be small, approximately $\le 10\degr$ \citep{JMJ}. In \textsc{ishem}, the value of $\phi$ has a strong influence on the flux normalisation and frequency of the spectral break (see Figure \ref{para}, top right panel). Similar to the effect of a varying inclination angle, a high opening angle shifts the final spectra towards lower frequencies and lower fluxes since a wider jet leads to a weaker magnetic field and, thus, to fewer energetic synchrotron photons. \\
 \indent \textcolor{black}{In this work, the jet power is left almost completely free (see Section \ref{powint}) as it may have varied a lot between the five observations of MAXI J1836-194 as discussed in \cite{RUSSA}. This parameter also has a strong influence on the resulting fluxes and on the break frequency (Figure \ref{para}, middle left panel). For a given index of the power-law distribution of electrons, a more powerful jet makes the emission mechanism more efficient resulting in more numerous and more energetic synchrotron photons.}\\ 
\indent Regarding the ejecta itself we used the conclusions derived in \cite{M2014}, namely, we assumed a relativistic flow by setting the adiabatic index to 4/3 and we used an initial volume filling factor of f$_{\text{v}}$=0.7. The shells of matter are ejected with a radius equivalent to 10 gravitational radii\footnote{\textcolor{black}{Typical dimension of the region of the accretion flow where a large part of the accretion power is dissipated in the hard state \citep{TINTIN}. Its impact here is negligible.}}. They are launched with a constant mass and are allowed to compress or expand. We chose to exclusively accelerate the electrons and we only take into account the energy losses due to radial expansion.\\ 
\indent The bulk Lorentz factors of compact jets in XRBs are very poorly constrained. Despite our lack of information on their exact values, it is commonly assumed that $\Gamma_{\text{XRBs}}$ are smaller than in active galactic nuclei $\Gamma_{\text{AGNs}}$$\sim10$. \cite{inf2} determined $\Gamma_{\text{XRBs}}$ $\leq$ 2 using the L$_{\text{X}}\propto$ L$_{\text{R}}^{0.7}$ correlation. However, it was later shown by \cite{heinz} that this correlation does not exclude high values of $\Gamma_{\text{XRBs}}$ and that XRBs are clearly capable of producing jets with Lorentz factors$\sim$ 10 \citep{JMJ}\footnote{The lack of strong constraints on XRBs Lorentz factors is also highlighted in the case of Cyg X-1 for which \cite{gleiss} and \cite{ZZZ} found different values of $\Gamma_{\text{av}}$ with, respectively, radio timing and a model of the jet anisotropy.}. \textcolor{black}{More recently, \cite{LORUP2} even found the exact opposite constraint in the case of GX339-4 where the jet Lorentz factor was constrained to be>2}. Hence, we considered a range of $\Gamma_{\text{av}}$$\sim$ 1-10. This parameter had unquestionably the strongest influence in our simulations (see Eq.\eqref{flux}, Eq.\eqref{freq} and the middle right panel in Figure \ref{para}). Indeed, a small increase of $\Gamma_{\text{av}}$ in the model moves the peak of the spectrum significantly towards lower frequencies and also causes the decrease of the jet flux. This is due to the fact that shells of matter with higher $\Gamma_{\text{av}}$ have a velocity closer to the speed of light. At higher $\Gamma_{\text{av}}$ the difference in shell velocities are smaller even if the difference in Lorentz factor is large. Therefore, it takes longer for the shells to catch up with each other and collisions occur at larger distances in the jet, in a larger region with weaker magnetic fields.\\
\indent The limits on the electron distribution have an impact on the jet power since they determine the properties of the population of electrons. For the same amount of kinetic energy transferred to the lepton distribution in shocks, a higher $\gamma_{\text{min}}$ increases both the number of very energetic particles and the average energy of the leptons leading to stronger and less absorbed emission (see Figure \ref{para}, bottom left panel). \textcolor{black}{In our model, the lower limit was initially set to $\gamma_{\text{min}}$=1 (electrons at rest) and the upper limit was frozen at $\gamma_{\text{max}}$=10$^{6}$ (typical value for XRB, \cite{M2014}). The electron distribution index, p, defines the slope of the optically thin part of the synchrotron spectrum (N$_{\text{e-}}$(E)$ \propto$ E$^{\text{-p}}$; Figure \ref{para}, bottom right panel). Both shock acceleration theory \citep{Krym,Bell} and observations of GX 339-4 \citep{indexindex,Drap} suggest p$\sim2.5$\footnote{Due to the high values of p>2, $\gamma_{\text{max}}$ has not a significant impact on the number of particles here.}.} 
\subsection{Fitting}
The fits to the multi-wavelength data were performed in four steps. The first step of the process involves computing, for each observation, a synthetic SED using the associated PSD and for a given set of parameters. \textcolor{black}{In a second step, the resulting spectrum is used as input in the X-Ray Spectral Fitting Package \citep[XSPEC ;][version 12.9.1p]{ARNO} via a local model, \textbf{ish}, with two independent parameters: the break frequency and break flux. To fit the broadband spectrum up to the X-rays, we associate the \textbf{ish} jet model with the irradiated disc model \textbf{diskir} \citep{diskir} to take into account the accretion disc contribution. This model fits the disc spectral signature as a standard \textcolor{black}{disc} blackbody and includes a Comptonisation component in order to fit the hard X-ray power law observed in XRBs. Moreover, \textbf{diskir} accounts for the irradiation of the inner and outer discs preventing the underestimation of the inner disc radius and fitting the reprocessed X-ray photons in the optical-UV band \citep{diskir}. The estimation of the jet contribution at high energies is done by extrapolating the optically thin part of jet spectra using power laws with identical slopes and by defining a high energy exponential cut-off in the synchrotron emission. We chose to set the cut-off energy to 20 eV (UV), as in \cite{RUSSA} due to lack of constraints. We also take into account the IR-optical and X-ray interstellar absorptions respectively with the \textbf{redden} model \citep{redden} and the \textbf{tbabs} model \citep{tbabs}. In the third step, once the appropriate shift in normalisation and frequency is found, we use Eq.\ref{flux} and Eq.\ref{freq} to determine a new set of physical parameters that would shift the initial SED by the required amount. There are seven parameters appearing in Eq.\ref{flux} an Eq.\ref{freq}, namely the distance to the source, the orbital inclination with respect to the line of sight, the jet power, the opening angle, the mean Lorentz factor, the electron distribution index and the lower limit of the electron distribution.} A large number of solutions/combinations exist causing a large degeneracy in the \textsc{ishem} main parameters. Thus, we freeze five parameters to reasonable values and we solve the equations for only two variables (see Section \ref{hivps}). We experimented with different parameters and found that the jet power and the jet opening angle produced the required changes in the SEDs for reasonable parameter ranges. In the last step, when a set of parameters is identified, we use \textsc{ishem} to compute the corresponding spectrum and compare it with the data. In general, the predictions of the analytic model are in good agreement with the simulations. It is important to notice that if no reasonable fit is found by shifting the initial synthetic SED, then the model is simply unable to reproduce the observations for any set of parameters since the shape of the SEDs are essentially defined by the power spectra of the input fluctuations.
\begin{figure*}
\centering
\includegraphics[scale=0.54]{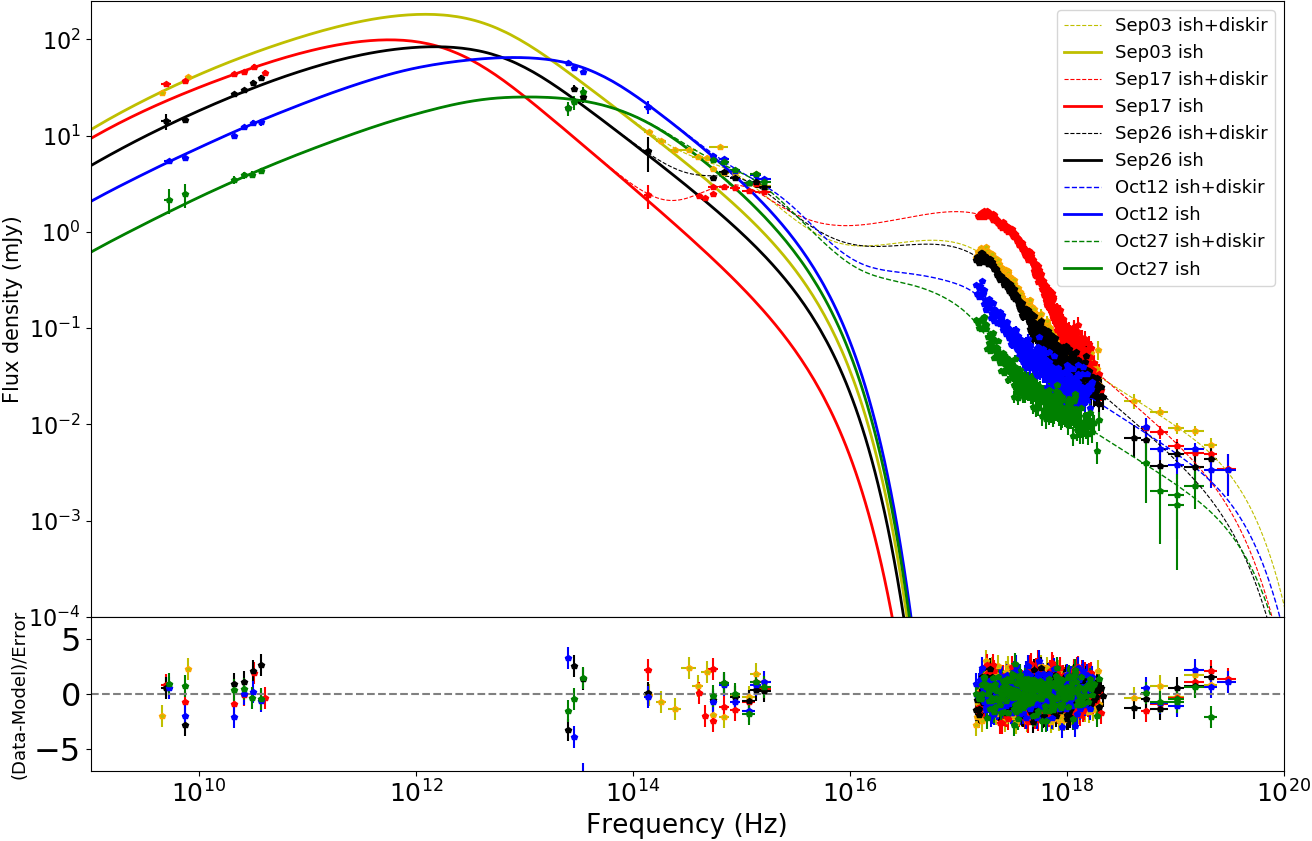}
\caption{ \textcolor{black}{[Upper panel] The best-fit spectral energy distributions determined with XSPEC for five of the multi-wavelength observational epochs of MAXI J1836-194. Solid lines represent the jet contributions obtained with the \textbf{ish} model and extrapolated up to 20 eV. Dashed lines represent the \textbf{ish+diskir} models used to account for the accretion disc contribution. [Lower panel] Fit residuals obtained with XSPEC (in terms of sigmas with error bars of size one, \textbf{delchi} plots). }}
\label{sed}
\end{figure*}
\begin{figure*}
\centering
 \includegraphics[scale=0.45]{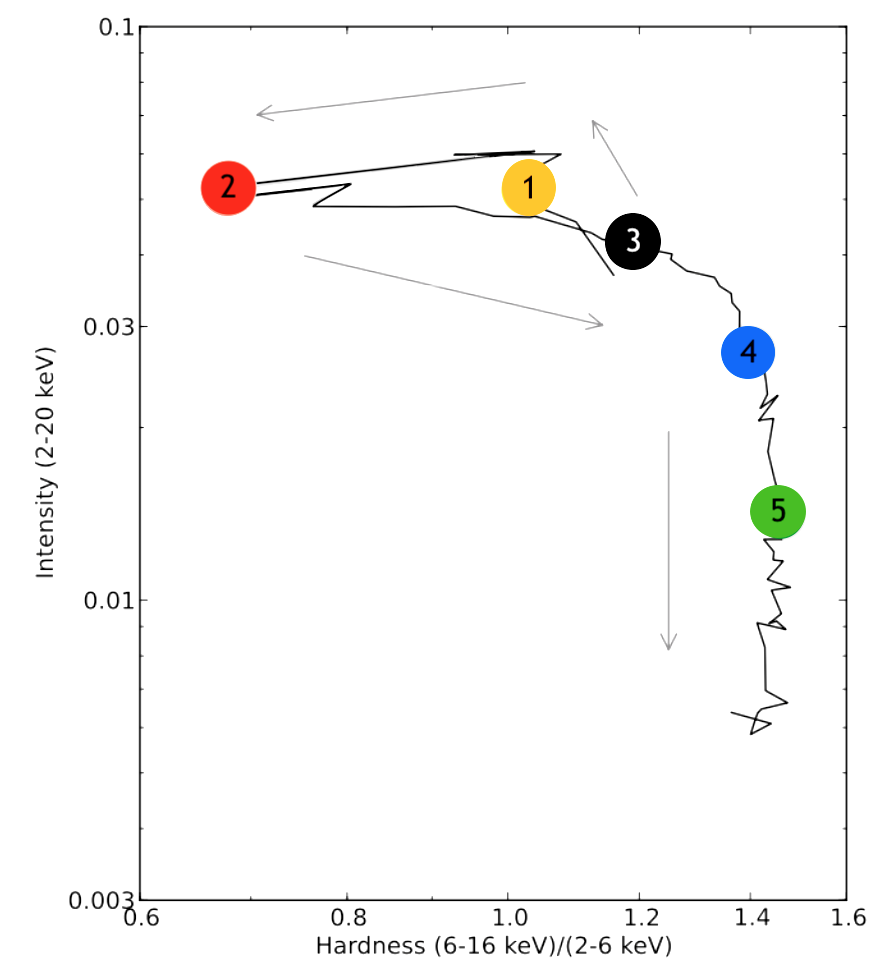}
\caption{Hardness-Intensity diagram of the 2011 outburst of MAXI J1836-194 \protect \citep[adapted from][]{DATADMRUSS}. 1:Sep 03, 2:Sep 17, 3:Sep 26, 4:Oct 12 \& 5:Oct 27. The coloured circles use the same colour code as in Fig.\ref{Tux}, Fig.\ref{para} and Fig.\ref{sed}. The black line shows the temporal evolution in the direction of the arrows.}
 \label{vc}
 \end{figure*}
\section{Results and discussion} %%%%%%% HERE SECTION THREE %%%%%%%%%%%%%%%%%%%%%%%%%%%%%%%%%%%%%%%%%%
\label{hivps}
\subsection{The minimal approach} 
As mentioned earlier, due to parameter degeneracies, there is a large number of parameter combinations that allow us to reproduce the observed evolutions of the SEDs shown in Figure \ref{sed}. Our approach was to try to explain this sequence by varying the minimum number of parameters across the five observations. As the mass of the central black, the distance to the source and the inclination are not likely to vary significantly over two months\footnote{Inclination angles might actually change through jet precession. Here we assume no precession.}, these parameters were kept fixed. Our first attempt aimed at reproducing the observed jet evolution with only one variable parameter. Unfortunately, explaining the sequence with only one variable parameter (e.g. varying only the jet power, $\Gamma_{\text{av}}$, or the jet opening angle) is not possible because it leads to simultaneous changes in both the flux normalisation and the spectral break frequency, both increasing or both decreasing (see Figure \ref{para}). Instead, Figure \ref{sed} shows that we need the jet break frequency to increase when the flux decreases to follow the observed jet evolution during the outburst. This implies that we need to vary at least two parameters in order to reproduce the observed sequence. Since the jet power and the mean Lorentz factor are expected to significantly change during the outburst due to the variation of the accretion rate, we chose to try to reproduce the five dates of the outburst by varying P and $\Gamma_{\text{av}}$.\\
\indent \textcolor{black}{We chose to set the mass of the central black hole and the distance to the source to reasonable values of their acceptable intervals, namely M=10.3 M$_{\odot}$ and D= 5kpc. These values were chosen during preliminary tests as they provided acceptable fit to the data. A wider range of masses and distances is explored in Section \ref{paraspace}. Using observational constraints on the width of the H$_{\alpha}$ emission line together with estimate of the disc size obtained from spectral fit to the SED, \cite{RUSSB} derived a relation between the mass, distance and inclination of the source. With the mass and distance chosen above we used this relation to fix the inclination angle to $5.5\degr$. These three parameters all respect the constraints established in Section \ref{paraspace}. We obtained an electron distribution index of 2.7 by fitting the optically thin part of the observations, however a somewhat steeper electron index of 2.9 was required to fit the data set of September 17. \textcolor{black}{We chose to freeze the jet opening angle at $\phi=1\degr$, consistent with the <2$\degr$ upper limit reported for Cyg X-1 jet in \cite{stirling}.}}
\subsection{Model spectra}
\label{troisdeux}
Figure \ref{sed} compares our best fit model SEDs to the observed data. As can be seen on this figure, the assumption that the jet fluctuations are driven by the X-ray PSDs leads to SED shapes that are very close to the observed ones. \textcolor{black}{With only one exception (September 17), the simulated SEDs are compatible with the source radio spectra. They only depart from the observations in the NIR where the influence of the accretion disc becomes significant (particularly on September 17 and on September 26 where the source is in the HIMS). The discrepancy between model and data in the radio band could stem from our poor knowledge of the X-ray power spectra. Indeed, in our model the shape of the radio part of the spectrum directly depends on the low frequencies of the PSD of the fluctuations where we do not have direct observation of the X-ray PSDs. The predicted radio flux is sensitive to our assumption of a flat noise extrapolation at frequencies below 1/512 Hz. Also, the non simultaneity and the radio variability \citep{gamup} during the measurement period are possibly substantial sources of error.\\
\indent When the source is in the hard state, we observe a brightening in the IR while the radio flux drops significantly. This causes the jet break to shift towards higher frequencies. On September 03 and 26 the break is around $10^{12}$Hz while at the hardest epochs, on October 12 and 27, the break is near $10^{13}$Hz. This is in qualitative agreement with the results of \cite{DATADMRUSS,RUSSA} especially in the hard state. 
\begin{figure}
\centering
\includegraphics[scale=0.35]{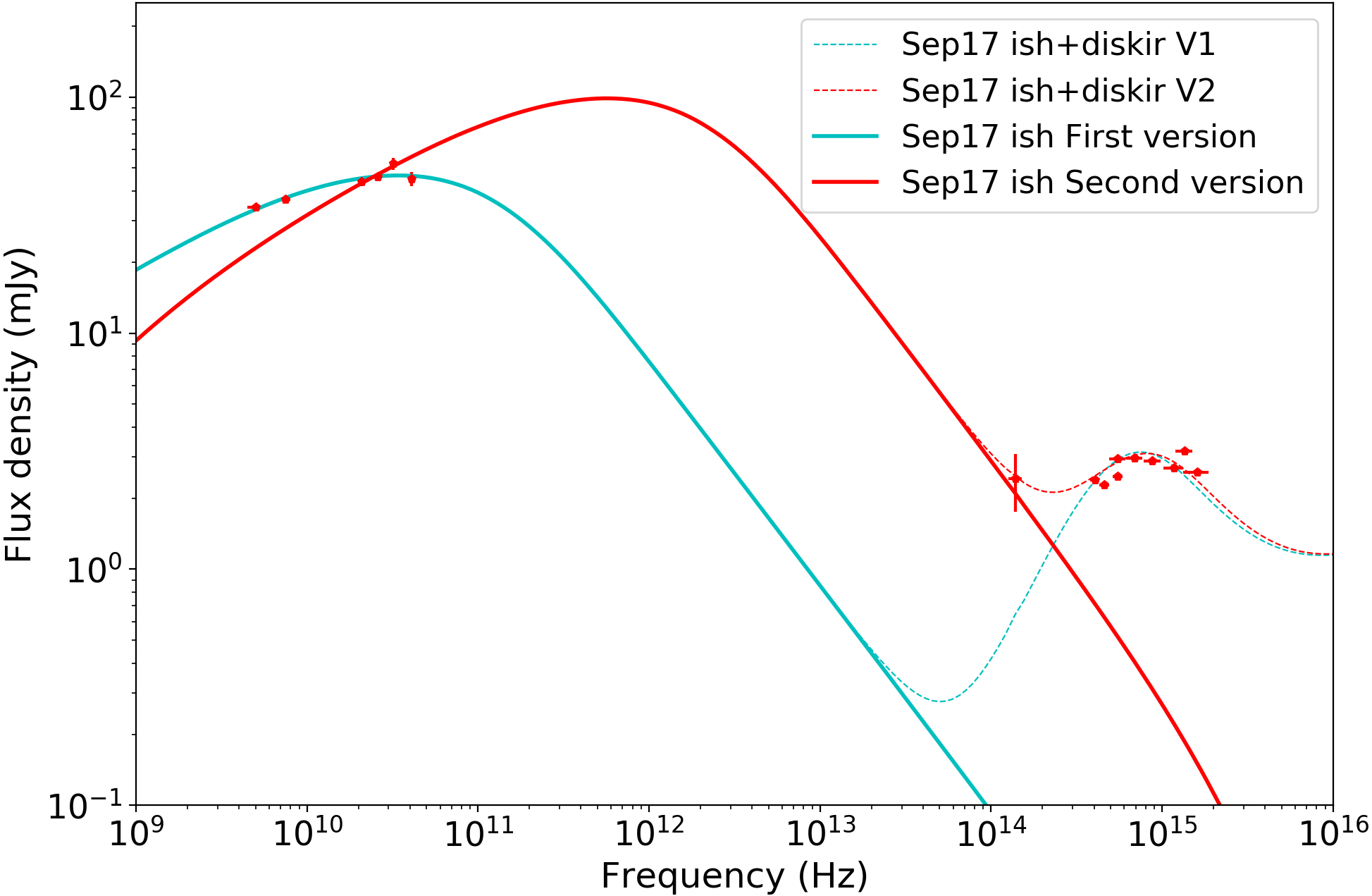}
\caption{ \textcolor{black}{The two versions of the September 17 spectral energy distribution determined with XSPEC (V1 on the left, V2 on the right). Solid lines represent the jet contributions obtained with the \textbf{ish} model. Dashed lines represent the \textbf{ish+diskir} models. See Section \ref{troisdeux}.}}
\label{sed2}
\end{figure}
The September 17 epoch could be modelled by two different jet SEDs (see Figure \ref{sed2}, described as first and second version) that are statistically equivalent but that stand out from each other by the way they pass through the data points. The first version represents the best fit obtained with XSPEC when using all the data. It goes through every radio point but does not go through the NIR point causing the SED to peak at low frequency, near 2x10$^{10}$Hz. In the second version, we ignore the first two radio points to fit the NIR data point leading to a frequency peak situated near $10^{12}$Hz. Since the first version of the fit required much more extreme physical parameters in order to produce the spectral turnover at cm wavelengths, notably in terms of mean Lorentz factor (where $\Gamma_{\text{av}}$>30 was required, see Figure \ref{para}), we decided to focus only on the second version of the fit.}
\subsection{The minimal scenario}
\label{powint}
\indent \textcolor{black}{The \textbf{diskir} parameters \textcolor{black}{and reduced chi-squares associated with our best fits are listed in Table \ref{my-label}. We obtained reduced chi-squares ranging from $\chi^{2}$=0.82 to $\chi^{2}$=1.27 that are practically equivalent to the ones obtained in \cite{RUSSA}. Although these values are good, it has to be noted that the goodness-of-fit measures are essentially dominated by the X-ray data (see lower panel of Figure \ref{sed} which shows the fit residuals obtained for the five dates of the study.)} The \textbf{diskir} parameters also appear to be nearly identical to the ones obtained in \cite{RUSSA}, notably the temperatures of the inner radius of the accretion disc. On the other hand, we tend to find different values of the logrout parameter (ratio of the outer disc radius in terms of the inner disc radius in logarithmic scale), especially on October 12 and October 27. \\
\indent It is possible to estimate the mean value of the jet power for the five epochs of the outburst using the measured X-ray luminosities \citep[see][and references therein]{Drap}:
\begin{equation}
\label{pow}
\text{P}\approx \textcolor{black}{43.6} \left [\frac{L_{X_{2-10 \text{keV}}}}  {\text{L}_{\text{E}}}\right ]^{0.5}   \text{\%L$_{\text{E}}$}.
\end{equation}
These estimates are indicative only as we note that Eq.\eqref{pow} is based on the model of \cite{powermodel} involving several assumptions which are not necessarily true for MAXI J1836-194. Namely, this assumes that the accretion flow is radiatively inefficient, that the jet receives a constant fraction of the accretion power and that the jet power, and the X-ray luminosity are equal to each other at L$_\text{{X}}$$\sim20$\text{\%L$_{\text{E}}$}. With M=10.3 M$_{\odot}$ and D=5kpc, we computed the five 2-10keV X-ray luminosities using XSPEC (see Table \ref{my-label}) and ended up with a range of luminosities going from 0.05\%L$_{\text{E}}$ to 0.36\%L$_{\text{E}}$ which results in a 0.96-2.62\%L$_{\text{E}}$ interval for the jet power to compare our results with.}\\
\label{hives}
\begin{table*}
\centering
\caption{\textcolor{black}{Main parameters of the simulations along with the \textbf{diskir} parameters obtained with XSPEC for a black hole mass of 10.3M$_{\odot}$, a distance to the source D$=5$kpc, an inclination angle $\theta=5.5\degr$ and an opening angle $\phi=1\degr$. nH represents the X-ray absorption in terms of hydrogen column density, logrout represents the log$_{10}$ of the outer radius in terms of inner radius (minimum value when logrout=3) and K refers to the normalisation. The electron temperature is here frozen at 100keV, the IR/optical/UV extinction, E(B-V), at 0.53 and the jet high frequency cut-off is set to 20 eV due to lack of constraints \citep[same as][for the last two parameters]{RUSSA}. Jet powers for $\gamma_{\text{min}}$=1 were calculated with slightly higher $\Gamma_{\text{av}}\text{s}$. The L$_{\text{X$_{2-10 \text{keV}}$}}$ luminosities were computed with XSPEC and P Eq.\ref{pow} refers to the jet power estimations obtained for these luminosities using Eq.\ref{pow}. The second part of the table represents the reasonable mean Lorentz factors and jet powers we obtained studying the parameter space (see Section \ref{paraspace}).}}
\label{my-label}
\vspace{1em}
\begin{tabular}{cccccc}

\hline
                                 & September 03 & September 17 & September 26 & October 12 & October 27 \\ \hline
Spectral state                   & Hard         & HIMS         & HIMS         & Hard       & Hard       \\
$\Gamma_{\text{av}}$                                   & 10.8         & 17           & 7.7          & 2.2        & 1.05       \\
P (L$_{\text{E}}$) for $\gamma_{\text{min}}$=8                      & 0.135        & 0.245        & 0.042        & 0.0135     & 0.003      \\
P (L$_{\text{E}}$) for $\gamma_{\text{min}}$=1                          & 0.324             &0.758              &0.099              & 0.03           &0.007            \\
p                                & 2.7          & 2.9          & 2.7          & 2.7        & 2.7        \\
\vspace{0.4em}
nH (x10$^{22}$cm$^{-2}$) & 0.197$^{+0.017}_{-0.016}$        & 0.286$^{+0.008}_{-0.008}$         & 0.231$^{+0.025}_{-0.023}$        & 0.310$^{+0.016}_{-0.016}$       & 0.394$^{+0.143}_{-0.052}$       \\
\vspace{0.4em}
kT\_disk (keV)                        & 0.239$^{+0.007}_{-0.007}$        & 0.429$^{+0.006}_{-0.004}$         & 0.232$^{+0.011}_{-0.012}$         & 0.104$^{+0.006}_{-0.007}$       & 0.103$^{+0.026}_{-0.024}$       \\
\vspace{0.4em}
Power-law index                  & 1.735$^{+0.037}_{-0.034}$        & 1.957$^{+0.062}_{-0.054}$         & 1.961$^{+0.358}_{-0.312}$         & 1.743$^{+0.041}_{-0.038}$       & 1.738$^{+0.081}_{-0.053}$       \\
\vspace{0.4em}
logrout                          & 4.085$^{+0.224}_{-0.208}$        & 4.209 $^{+0.101}_{-0.109}$        & 3.938 $^{+0.163}_{-0.156}$        & 3.000$^{+0.338}$     & 3.332$^{+0.501}_{-0.332}$       \\
\vspace{0.4em}
K (x10$^{3}$)       & 20.511$^{+3.778}_{-3.142}$       & 7.326$^{+0.633}_{-0.574}$         & 19.810$^{+7.731}_{-4.971}$        & 92.414$^{+25.959}_{-18.475}$      & 92.037$^{+129.925}_{-33.629}$     \\\hline
%\vspace{0.4em}

\text{$\chi$}$^{2}$/d.o.f.                      & 529.31/444   & 581.25/490   & 389.62/478   & 419.58/331 & 264.17/275\\ \hline 
L$_{\text{X$_{2-10 \text{keV}}$}}$ (L$_{\text{E}}$)& 0.0021 &0.0036  & 0.0016 & 0.001 &0.0005  \\
P Eq.\ref{pow} (L$_{\text{E}}$)& 0.0198 & 0.0262& 0.0175 &0.0141 & 0.0096\\ \hline \hline
$\Gamma_{\text{avF}}$   &10.45 &16 & 7.55& 2.15&1.045\\
P$_{\text{F}}$&0.047 &0.039 &0.0212&0.0114&0.0026
\end{tabular}
\end{table*}
\indent {\textcolor{black}{We obtain five parameter sets in which the jet power and the mean Lorentz factor both increase with the source luminosity. The associated parameters and the goodness-of-fit are listed in Table \ref{my-label}. Keeping the minimum energy of the electrons $\gamma_{\text{min}}$ frozen to unity implied jet kinetic powers that could be much larger than the estimates of $\sim$2.62\%L$_{\text{E}}$ provided by Eq.\ref{pow}. We thus used higher values of $\gamma_{\text{min}}$ in order to lower P (see Table \ref{my-label}). However, setting $\gamma_{\text{min}}$ too high can remove the particles that produce the synchrotron emission at frequencies of interest (typically when the leptons at energy $\gamma_{\text{min}}$ emit photons with frequency higher than the synchrotron self-absorption turnover frequency). We checked that the shape of the predicted SEDs is not affected as long as $\gamma_{\text{min}}$ is below 8.} If $\gamma_{\text{min}}$ is set to 8 rather than 1, the required jet power is divided by a factor of 2-3 and remains below {\textcolor{black}{25}\%L$_{\text{E}}$. The jet power values obtained for $\gamma_{\text{min}}$=8 and $\gamma_{\text{min}}$=1 are listed on the third and fourth row of Table \ref{my-label}. In the end, we obtained jet powers ranging from {\textcolor{black}{0.3 to 24.5}\%L$_{\text{E}}$ and $\Gamma_{\text{av}}$ values ranging from {\textcolor{black}{1.05 to 17} for the five epochs of the study (P=24.5\%L$_{\text{E}}$ and $\Gamma_{\text{av}}$={\textcolor{black}{17} being reached in the HIMS on September 17, see Fig.\ref{vc}). The maximum jet power value, P= 24.5\%L$_{\text{E}}$, still represents more or less 70 times the observed X-ray luminosity and consequently requires 70 times more accretion power put into the jets than radiated away in the disc. This raises some issues for accretion disc models since an insufficiently heated disc results in a X-ray spectrum dominated by thermal emission which is inconsistent with the power-law shaped spectrum we observe. Moreover, the jet power in the hard state (HIMS here) should be released as radiative power in the soft state where the jet is quenched and the accretion flow radiatively efficient. If the jet power is P$\approx$70L$_{\text{X}}$ in the hard state, it implies that the X-ray luminosity should sharply increase by a factor of 70 during state transitions. However, such jumps of X-ray luminosity during transitions have never been observed.\\
\indent We failed to find any reasonable parameter set for the three epochs of September by varying together only the jet power and the jet opening angle as they would be unable to sufficiently shift the spectra without using extreme values, namely, super-Eddington luminosities and $\phi$>60\degr. We were also unable to reproduce the data with lower jet powers or lower mean Lorentz factors, because it either required unrealistic values for the other parameters or increased the number of free parameters. 
\section{Parameter space} %%%%%%% HERE SECTION FOUR %%%%%%%%%%%%%%%%%%%%%%%%%%%%%%%%%%%%%%%%%%
\label{paraspace}
We have shown that jet emission powered by internal shocks driven by the accretion flow variability can provide a good description of the broadband data of MAXI J1836-194 (Section \ref{hivps}). However, due to the large parameter degeneracy it is not possible to simultaneously constrain all of the model parameters. Instead, only the jet power and average jet Lorentz factor were allowed to vary while fitting the five data sets, where the other parameters were kept fixed at reasonable values. This notably led to jet power values that exceed the estimation of Eq.\ref{pow}.\\
\indent  \textcolor{black}{In this section, we explore in more details the parameter space allowed by the data by changing the values of opening angle, mass, distance and inclination angle. Note that those parameters remain fixed from one observation to another. The main objective being to find fitting parameter sets using smaller $\Gamma_{\text{av}}$ and jet power values.}
\subsection{Opening angle}
 {\textcolor{black}{We first examine how the jet opening angle and the jet power vary with the mean Lorentz factor, in particular for the first three observations of the outburst. We seek to reduce the interval of $\Gamma_{\text{av}}$ (1.05 to 17) obtained in Table \ref{my-label} down to a more reasonable range, namely with a maximum $\Gamma_{\text{av}}$$\lesssim10$. Figure \ref{jetopen} show the required values for $\phi$ and P when $\Gamma_{\text{av}}$ is varying from 1.1 to 20. All the other parameters were kept at the same values as in Table \ref{my-label}. In this figure, each point corresponds to a parameter set that fits the multi-wavelength observations. We show that the decrease of the mean Lorentz factor directly implies to the increase of the opening angle to compensate for the frequency shift (see Figure \ref{para}) and inevitably to the increase of jet powers in order to maintain the correct fluxes. Allowing the opening angle to increase up to the upper limit of its acceptable range, namely about 10$\degr$, it is possible to reduce the $\Gamma_{\text{av}}$ interval to 1.05-9.1 (see Table \ref{fig4}). However, since our main goal is also to approach the 0.96-2.62\%L$_{\text{E}}$ jet power interval, such an increase in the values of the jet power turns out to be a very undesirable effect.} \\
\begin{table}
\centering
\caption{\textcolor{black}{Possible reduction of $\Gamma_{\text{av}}$ values using wider opening angles and more powerful jets. \textcolor{black}{Values between square brackets} refer to the parameters displayed in Table \ref{my-label}.}}
\label{fig4}
\begin{tabular}{c|cccccc}
             & \multicolumn{2}{c}{$\Gamma_{\text{av}}$} & \multicolumn{2}{c}{$\phi$} & \multicolumn{2}{c}{P (L$_{\text{E}}$)}  \\ \hline
September 03 & 5.7   & [10.8]  & 9.77$\degr$    & [1$\degr$]    & 0.273 & [0.135] \\
September 17 & 9.1   & [17]    & 9.91$\degr$    & [1$\degr$]    & 0.34  & [0.245] \\
September 26 & 4     & [7.7]   & 9.88$\degr$    & [1$\degr$]    & 0.112 & [0.042]
\end{tabular}
\end{table}
\begin{figure}
\centering
\includegraphics[scale=0.34]{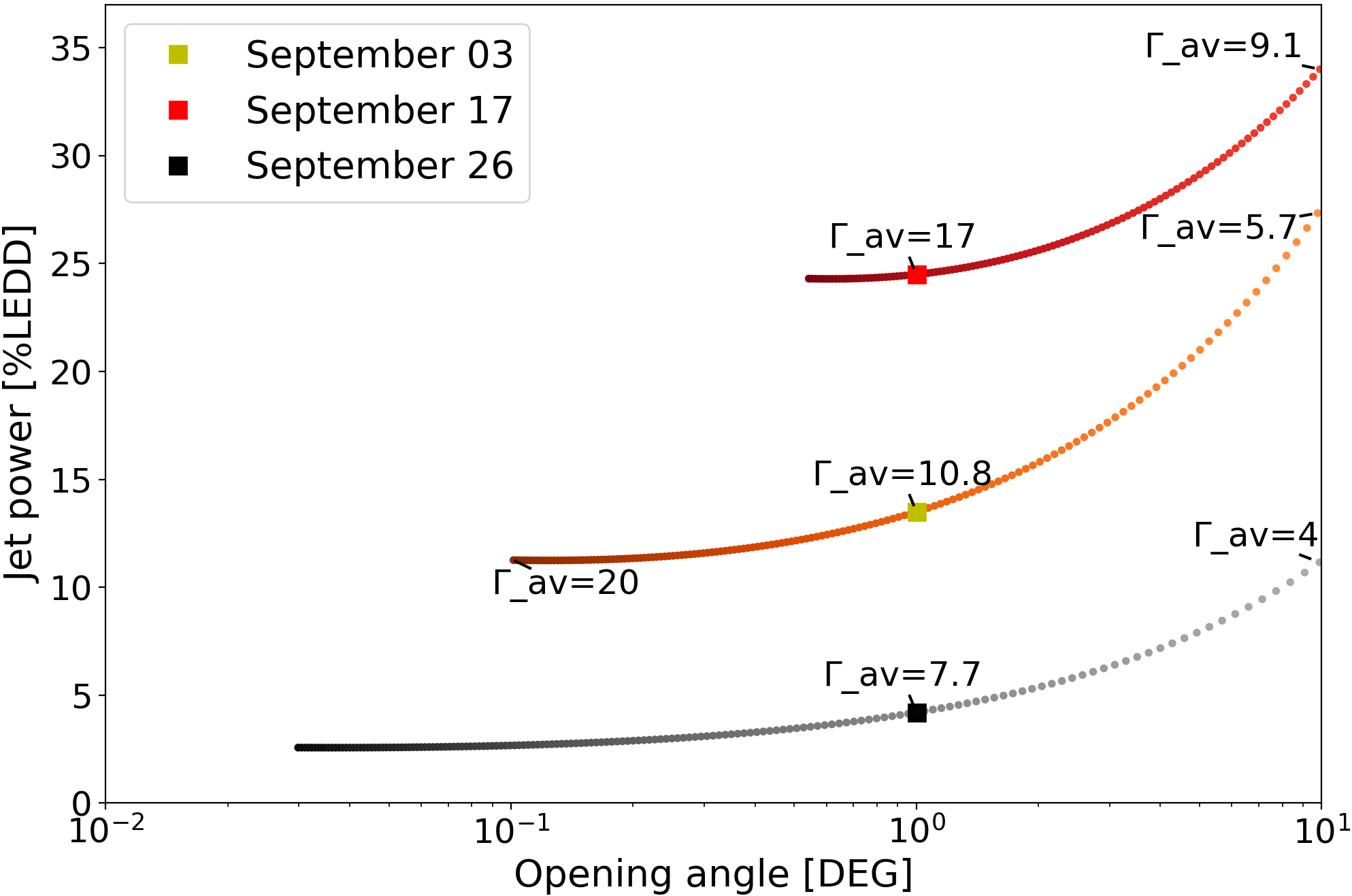}
\caption{\textcolor{black}{Evolution of the jet power and of the jet opening angle as a function of the mean Lorentz factor for the three first dates of the outburst. Each point represents a parameter set that matches the observations. The colour gradients illustrate the increase of $\Gamma_{\text{av}}$. The squares depict the parameters we obtained in section \ref{hivps}.}}
\label{jetopen}
\end{figure}
\begin{figure*}
\centering
\includegraphics[scale=0.70]{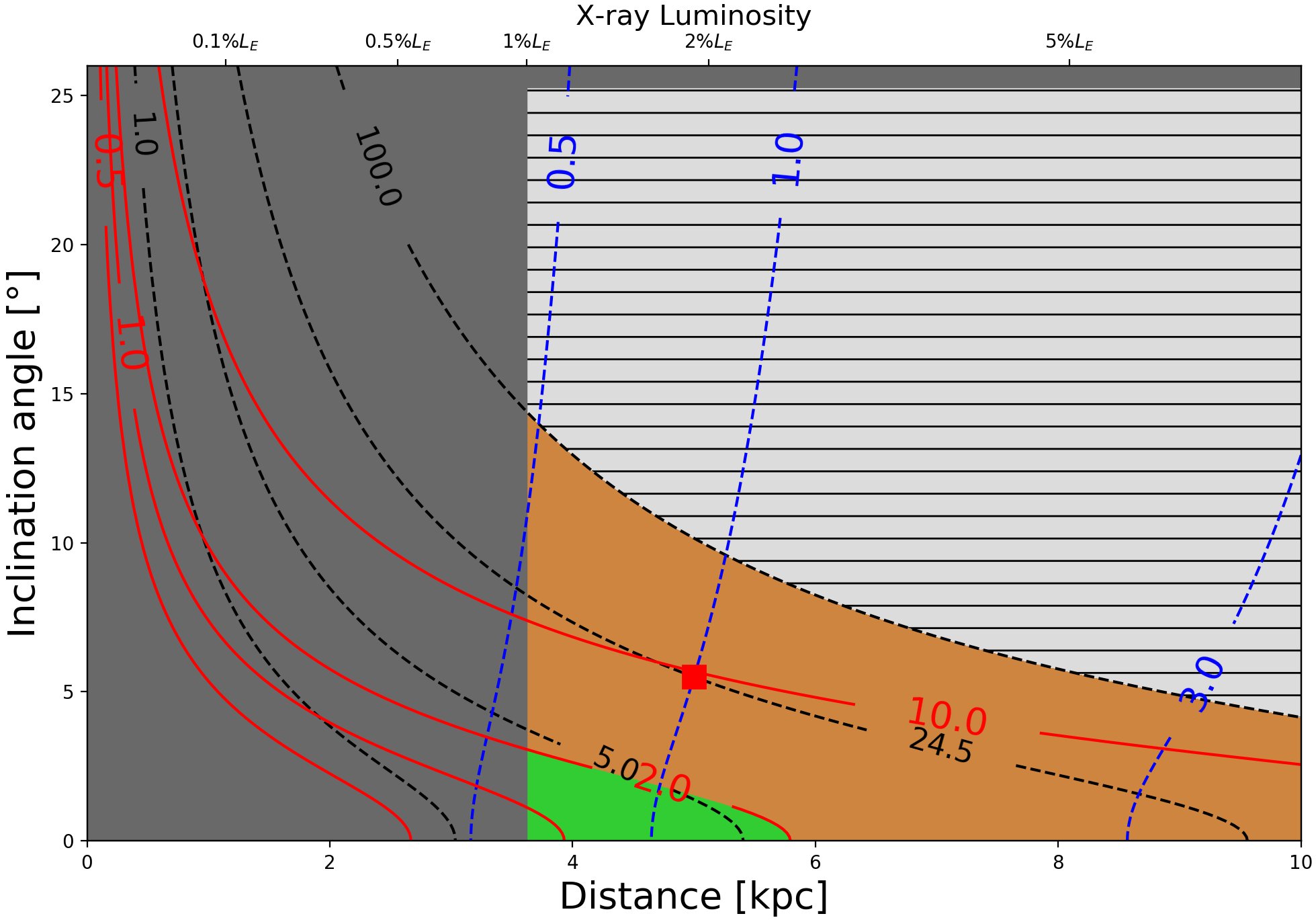}
\caption{\textcolor{black}{Parameter map showing the evolution of the jet power and of the jet opening angle as a function of the distance and the inclination angle for M=10.3 M$_{\odot}$ on the softest dataset of the outburst (September 17). Black dashed contours represent jet power values in Eddington units, vertical blue dashed contours represent opening angle values in degrees and red solid contours correspond to the ratio between P and P$_{\text{Eq}.\ref{pow}}$. The hatched area represents the super-Eddington domain \textcolor{black}{(P>100\%L$_{\text{E}}$)} while the dark grey zones depict the excluded values of distance and inclination angle obtained from equation \ref{cinqk} and \ref{onze}. The green area corresponds to the jet power values that follows the Eq.\ref{pow} estimation and the \textcolor{black}{orange} area corresponds to \textcolor{black}{jet power values that are higher than 2*P$_{\text{Eq}.\ref{pow}}$}. The red square represents the parameters from Table \ref{my-label}.}}
\label{contour}
\end{figure*}
\begin{figure*}
\centering
\includegraphics[scale=0.70]{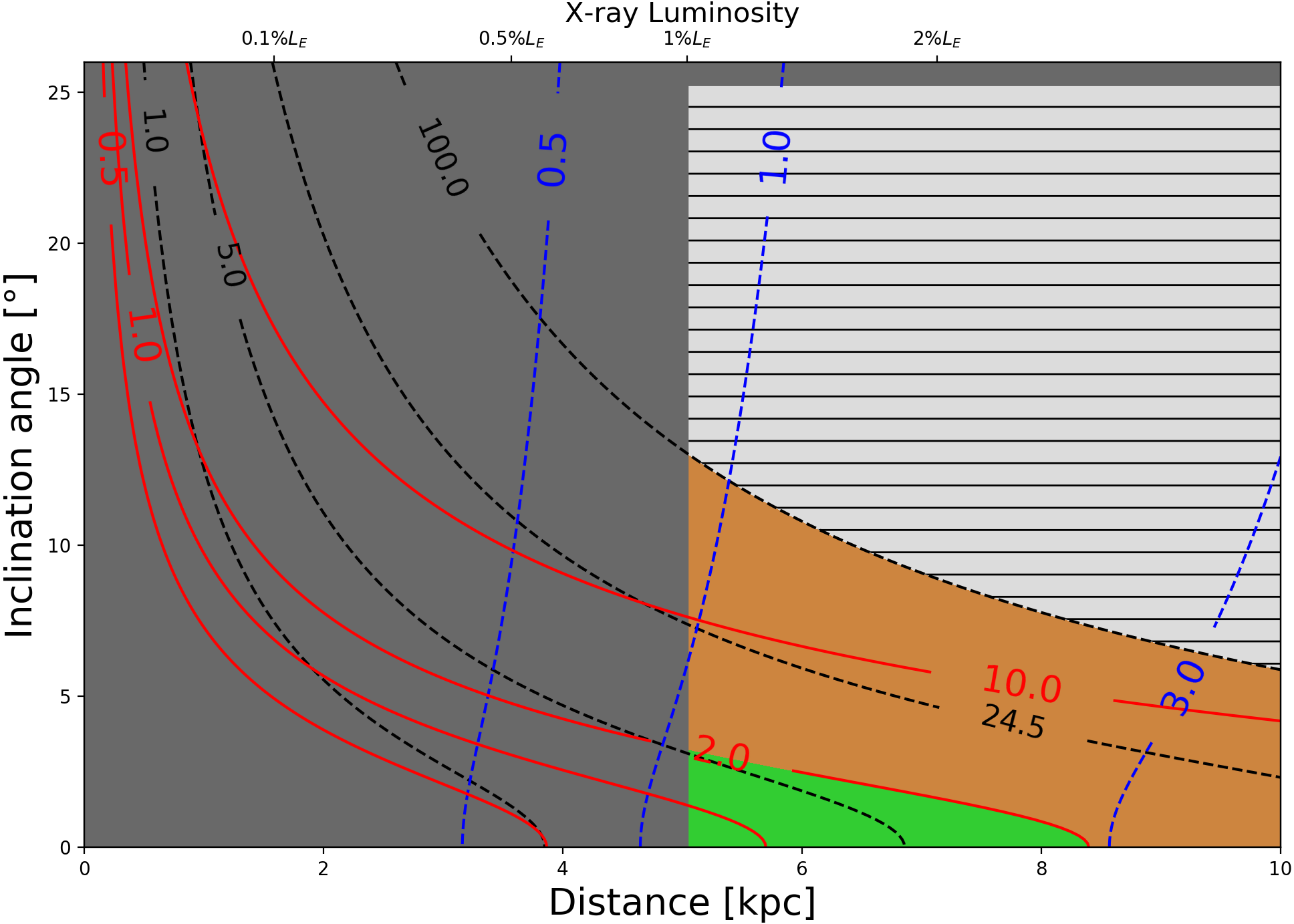}
\caption{\textcolor{black}{Parameter map showing the evolution of the jet power and of the jet opening angle as a function of the distance and the inclination angle for M=20 M$_{\odot}$ on the softest dataset of the outburst (September 17). The colours and lines are the same as Figure \ref{contour}.}}
\label{contour20}
\end{figure*}
\subsection{Mass, distance and inclination angle}
\textcolor{black}{We now seek to reduce our values of jet power by studying how the jet opening angle and the jet power values change when modifying the last three remaining main parameters of the simulation which are the mass of the black hole, the distance to the source and the inclination angle. We first derive below some observational constraints on these parameters in order to reduce the search space.
%We decide to study the influence of these three parameters together since we can obtain relation linking them using the \textbf{diskir} parameters we obtained in Section \ref{powint}..
% In the following, we perform our own estimations of MAXI J1836-194 inclination angle, distance and mass with the help of the \textbf{diskir} parameters we obtained in Section \ref{powint}.
}
\subsubsection{Inclination angle}
\textcolor{black}{The inclination angle of MAXI J1836-194 can be directly estimated from the measure of the \textbf{diskir} logrout parameter (see \ref{powint} hereafter referred as y) following equation 1 of \cite{RUSSB}:}
\begin{equation}
\label{deux}
\textcolor{black}{\text{sin}\theta= \text{v$_{\text{out}}$} \sqrt{\frac{10^{\text{y}} \text{R}_{\text{in}}}{\text{GM}}}= \text{v$_{\text{out}}$} \sqrt{\frac {\text{R}_{\text{out}}}{\text{GM}}}}
\end{equation}
\textcolor{black}{where v$_{\text{out}}$ represents the rotational velocity of the outer disc, R$_{\text{in}}$ the physical inner disc radius and R$_{\text{out}}$ the physical outer disc radius. Assuming R$_{\text{in}}$=R$_\textsc{isco}$ on the softest date of the outburst as in \cite{RUSSA} \citep[see][]{millerisco,reisco}, with R$_\textsc{isco}$=6$\mu$GM/c$^{2}$ ($\mu$ depends on the black hole spin: $\mu$=1 for a Schwarzschild black hole and $\mu$=1/6 for an extreme Kerr black hole, we obtain:}
\begin{equation}
\label{cinq}
\textcolor{black}{\text{sin}\theta= \frac{v_{\text{out}}}{\text{c}}\sqrt{10^{\text{y}}6\mu}= \frac{v_{\text{out}}}{\text{c}}\sqrt{\text{R}_{\text{out}}/\text{R}_{\text{in}}6\mu}}
\end{equation}
\textcolor{black}{ \cite{RUSSB} found that the Keplerian velocity of the disc ring that gives the highest H$_{\alpha}$ contribution could be used as a good approximation of the velocity of the outer disc and thus estimated v$_{\text{H$_{\alpha}$}}$=v$_{\text{out}}$=130km.s$^{-1}$. However, since the disc ring is not necessarily at the outer edge of the accretion disc, we rather consider v$_{\text{H$_{\alpha}$}}$ as an upper limit, namely v$_{\text{H$_{\alpha}$}}$=130km.s$^{-1}$$\geq$v$_{\text{out}}$. Similarly, the Sep 17 dataset (softest of the data epochs studied, see Figure \ref{vc}) is in the hard-intermediate state, therefore the physical inner radius is most likely further from the ISCO. In the hard state the disc is likely truncated at about 50-100 gravitational radii \citep{trunk}, thus choosing an intermediate value of 10r$_{\text{g}}$ in the HIMS we have: R$_{\text{in}}$=$\text{f}\text{R}$$_\textsc{isco}$, with 1$\leq$f$\leq$10. Equation \ref{cinq} then simply becomes:}
\begin{equation}
\label{cinqk}
\textcolor{black}{\text{sin}\theta= \frac{v_{\text{out}}}{\text{c}}\sqrt{10^{\text{y}}6\text{f}\mu}= \frac{v_{\text{out}}}{\text{c}}\sqrt{\text{R}_{\text{out}}/\text{R}_{\text{in}}6\text{f}\mu}}
\end{equation}
\textcolor{black}{On this basis, we are only able to determine the upper limit of the inclination angle corresponding to the case where v$_{\text{out}}$=130km.s$^{-1}$ and f$\mu$=10 (R$_{\text{in}}$=10R$_\textsc{isco}$ for a Schwarzschild black hole). With y=4.209 (see Table \ref{my-label}), it leads to $\theta$$\leq$25.29$\degr$. }
\subsubsection{Mass-distance relations}
\textcolor{black}{Knowing the inclination angle, it is then possible to derive a relation between the mass of the black hole and the distance to the source from the measure of the \textbf{diskir} normalisation parameter K. It is defined as follows: K=(r$_{\text{in}}$)$^{2}$(10/D$_{\text{kpc}}$)$^{2}$cos$\theta$, where r$_{\text{in}}$ represents the apparent inner disc radius, related to the physical inner disc radius according to R$_{\text{in}}$$\approx$1.19r$_{\text{in}}$ \citep{shimura,kubota, soria}, and D$_{\text{kpc}}$ the distance in kiloparsec units. Assuming R$_{\text{in}}$=$\text{f}\text{R}$$_\textsc{isco}$, it becomes:}
\begin{equation}
\label{neuf}
\textcolor{black}{\frac{\text{M}}{\text{M}_\odot}=\sqrt{\frac{\text{K}}{\text{cos}\theta}}\ \frac{D_{\text{kpc}}1.19\text{c}^{2}} {60 \text{f}\mu \text{G}\text{M}_\odot}}
\end{equation}
\textcolor{black}{With K=7326 (see Table \ref{my-label}), it results in the following mass-distance relation:}
\begin{equation}
\label{dix}
\textcolor{black}{0.14\frac{\text{M}}{\text{M}_\odot} \leq \text{D}_{\text{kpc}} \leq 8.33 \frac{\text{M}}{\text{M}_\odot}}
\end{equation} 
\textcolor{black}{It is also possible to determine the mass-distance relation using the constraints on the luminosities required for the source to transit from the hard state to the HIMS (and vice versa). \textcolor{black}{\cite{dudu} showed that the hard to HIMS state transition in black hole binaries occurs for bolometric disc luminosities (disc + high energy power law\footnote{\textcolor{black}{In \cite{RUSSA} only the disc luminosity is taken into account, leading to a different mass-distance relation.}}) larger than 1\%L$_{\text{Edd}}$ (see Fig.10, lower panel)}. Therefore assuming that the softest point of MAXI J1836-194 outburst is >1\%L$_{\text{Edd}}$, we can write:}
\begin{equation}
\label{onze}
\textcolor{black}{\frac{\text{L}}{\text{L}_\text{Edd}}=\frac{\text{F} 4\pi D_{\text{cm}}^{2}}{1.26\ 10^{38}\frac{\text{M}}{\text{M}_\odot}}> 0.01}
\end{equation}
\textcolor{black}{with F the 0.1-100keV flux expressed in erg.cm$^{-2}$.s$^{-1}$ (calculated from the XSPEC models) and D$_{\text{cm}}$ the distance expressed in cm. Using F=8.28.10$^{-9}$erg.cm$^{-2}$.s$^{-1}$, we get D$^{2}$$_{\text{kpc}}$>1.27$\frac{\text{M}}{\text{M}_\odot}$ which is more constraining that the left-hand side of Eq.\ref{dix}.}
\subsubsection{Parameter map}
\textcolor{black}{To study the impact of the distance and of the inclination angle on the jet power, we create a parameter map (see Figure \ref{contour}) that shows the fitting values of jet power and opening angle calculated with equations \ref{flux} and \ref{freq} when modifying D and $\theta$ on the September 17 dataset. The main purpose is to identify, on the epoch requiring the highest jet power, an area of parameter space for which the jet power can be reduced down to the estimations obtained with equation \ref{pow}. In order to scan a large part of the parameter space, we use distances ranging from 3.62kpc (lower limit found using D$^{2}$$_{\text{kpc}}$>1.27$\frac{\text{M}}{\text{M}_\odot}$) to 10kpc and inclination angles ranging from 0$\degr$ to 25.29$\degr$, setting the black hole mass to 10.3 M$_{\odot}$. In Figure \ref{contour}, we show that the intervals of distance and inclination angle proposed in \cite{RUSSB, RUSSA} inevitably lead to jet powers that are at least two times higher than the estimations (\textcolor{black}{orange} area) or even super-Eddington (hatched area) for high D and high $\theta$. On the contrary, jet power values compatible with the estimations (green area) are obtained using low inclination angles and low distances (excluding the values that do not respect the constraints we found using equations \ref{cinq}, \ref{dix} and \ref{onze} in dark grey) that can be inferior to the lower limits of those intervals. Allowing  \textcolor{black}{-arbitrarily-} the jet power to attain twice the value of P$_{\text{Eq}.\ref{pow}}$, we can obtain reasonable parameter sets for distances ranging from 3.6kpc to 5.8kpc and inclination angles that are lower than 3$\degr$, with M=10.3M$_{\odot}$. Keeping the opening angle set to 1$\degr$ but decreasing the inclination angle value down to 1$\degr$ and the distance down to 4.7kpc, we were able to make a new tracking of the compact jet evolution during the 2011 outburst using reasonable jet power values (see the last two rows of Table \ref{my-label}). In this new parameter set, we obtained jet power values ranging from 0.26\%L$_{\text{E}}$ to 4.40\%L$_{\text{E}}$ and a mean Lorentz factor interval of 1.045-16. The jet evolution is also reproduced with $\Gamma_{\text{av}}$ and P increasing with the source luminosity, in the hard state at least, since the maximum jet power is no longer attained on September 17 but on September 03. \\
\indent In a toy model proposed in order to explain MAXI J1836-194 odd L$_{\text{X}}$$\propto$L$_{\text{R}}^{\sim1.8}$ behaviour \cite{gamup} suggested that the distance to the source should be superior to 8kpc. In this model, the authors investigated the possibility that variable relativistic beaming was responsible for the steep correlation. According to Figure 9 of \cite{gamup}, the variable Doppler boosting cannot account for the odd 1.8 correlation for distances lower than 8kpc since it would sometimes require boosting and deboosting at other times which is not expected with such very low inclination angles. In our case, we see that the maximum distances that can be reached for this specific black hole mass are below the 8kpc threshold and are paired with inclination angles that are lower than 2$\degr$. Interestingly we show, in Figure \ref{contour20}, that increasing the mass of the central black hole up to M=20M$_{\odot}$ can shift the very limited region of interest toward the high distances and high opening angles so that we can reach the 8-10kpc interval suggested by \cite{gamup}. However, such a scenario implies even lower inclination angles. In both scenarios, the use of low inclination angles makes MAXI J1836-194 a strong microblazar candidate.}
\section{Conclusion} %%%%%%% HERE THE CONCLUSION %%%%%%%%%%%%%%%%%%%%%%%%%%%%%%%%%%%%%%%%%% 
\label{dis}
In this paper, we have presented a method to fit the multi-wavelength emission of jets in microquasars. The apparent small angle between the compact jet of the black hole candidate MAXI J1836-194 and the line of sight, along with the availability of excellent multi-wavelength observations made this source an ideal candidate for our study. The main result of this work is that an internal shock model in which the shocks are driven by the accretion flow variability can successfully reproduce the SEDs of the compact jet of MAXI J1836-194 for five observational epochs during its 2011 outburst. Our model is able to produce the observed shift of the jet break \citep[e.g.][]{DATADMRUSS,RUSSA} in the SEDs by varying the jet bulk Lorentz factor. The variation of the jet break with the source hardness has been seen in other sources as well \citep[e.g. in MAXI J1659-152 where the break was at even lower frequencies when this source was even softer][]{VDH, karri}.\\
\indent We showed that the evolution of the jet through the hard and hard-intermediate state could not be fitted with only one variable parameter but with at least two parameters. We obtained consistent fits by varying the jet power together with the jet mean Lorentz factor, with both parameters increasing with the luminosity. This result corroborates the toy model suggested in \cite{gamup} to explain MAXI J1836-194 peculiar radio/X-ray correlation (L$_{\text{X}}\propto$ L$_{\text{R}}^{\sim1.8}$) with variable relativistic boosting requiring the jet mean Lorentz factor to increase with the source \textcolor{black}{luminosity} \citep[although, while their toy model could work for MAXI J1836-194, it could not for all other systems;][]{gamup}. The values of the jet bulk Lorentz factors in our minimal scenario are quite large with a maximum of \textcolor{black}{$\Gamma_{\text{av}}$=17} being reached on September 17 when the source had the lowest hardness (HIMS). For the hard state epochs, the values appear similar to those suggested by \cite{gamup}, $\Gamma_{\text{av}}\sim 1-2$. \cite{fender2} suggested that the jet bulk Lorentz factor increases as a black hole XRB softens over the hard to soft state transition (see their Figure 7, lower panel). The results here seem to support this scenario. As for the jet power values, they largely exceed the estimations obtained with equation \ref{pow} on the first three epochs. The maximum being also reached on September 17 with P= 24.5\%L$_{\text{E}}$. \\
\indent However, we showed that this scenario is not unique due to the degeneracy of the parameters of the \textsc{ishem} model. \textcolor{black}{By allowing a free jet opening angle, the interval of \textcolor{black}{mean Lorentz factors} could be 1.05-9.1 using wider and more powerful jets. Our in-depth exploration of the parameter space performed in order to reduce \textcolor{black}{the jet power interval} led to the determination of a very constrained region where the jet powers are comparable to the X-ray luminosities and in agreement with phenomenological estimates. Reaching it implies to use low distances, low opening angles and particularly low inclination angles. We managed to reduce the jet power interval down to 0.26-4.40\%L$_{\text{E}}$ with the maximum no longer being reached on September 17 but on September 03, most likely due to the steeper electron energy distribution found on September 17 associated to a very low inclination angle. This suggests that the increase of the jet power with the source luminosity occurs at least in the hard state. The maximum value of the mean Lorentz factor is reduced to $\Gamma_{\text{av}}$=16. Using a more massive central black hole, we also showed that it is possible to shift this limited region towards higher distances which appear to be required to explain the odd L$_{\text{X}}\propto$ L$_{\text{R}}^{\sim1.8}$ behaviour.}\\
\indent \textcolor{black}{Finally, we have determined a very narrow region of the parameter space in which we can use acceptable jet power values and parameters that respect all the physical constraints. To track the compact jet evolution through the 2011 outburst, we only need to vary two parameters: the jet power and the mean Lorentz factor. The latter increases with the source luminosity and decreases with the source hardness for the five epochs of the study with a maximum value of $\Gamma_{\text{av}}$=16. As for the jet power, it seems to follow this behaviour in the hard state at least. Obtaining reasonable jet powers at all epochs requires very small jet inclinations, of a few degrees at most. This confirms that MAXI J1836-194 could be a microblazar and explains the compact jet dominance up to the IR-optical. Future tests of the model against other black hole X-ray binaries data will help to establish whether our inferred evolution of jet Lorentz factor during the outburst can be generalised to other objects.
This kind of studies requires multi-wavelength monitoring throughout an outburst with good coverage not only in radio and X-rays but also in optical, IR, and sub-mm bands.}
%FUTURE PROSPECTS 
\section*{Acknowledgements}
This work is part of the CHAOS project ANR-12-BS05-0009 supported by the French Research National Agency (\url{http://www.chaos-project.fr}). It was made possible thanks to the support from PNHE in France and from the OCEVU Labex (ANR-11-LABX-0060) and the A*MIDEX project (ANR-11-IDEX-0001-02) funded by the Investissement d'Avenir French government program managed by ANR. This research has made use of data obtained from the High Energy Astrophysics Science Archive Research center (HEASARC), provided by NASA's Goddard Space Flight Center. SC and JR acknowledge financial support from the UnivEarthS Labex program of Sorbonne Paris Cité (ANR-10-LABX-0023 and ANR-11-IDEX-0005-02). TDR acknowledges support from the Netherlands Organisation for Scientific Research (NWO) Veni Fellowship, grant number 639.041.646.

%%%%%%%%%%%%%%%%%%%%%%%%%%%%%%%%%%%%%%%%%%%%%%%%%%

\bibliographystyle{mnras}
\bibliography{biblio} % if your bibtex file is called example.bib

%%%%%%%%%%%%%%%%%%%%%%%%%%%%%%%%%%%%%%%%%%%%%%%%%%

%%%%%%%%%%%%%%%%% APPENDICES %%%%%%%%%%%%%%%%%%%%%

%\appendix

%%%%%%%%%%%%%%%%%%%%%%%%%%%%%%%%%%%%%%%%%%%%%%%%%%
% Don't change these lines
\bsp	% typesetting comment
\label{lastpage}
\end{document}